\documentstyle[twoside,11pt]{article}
\normalsize
\begin{document}
\bibliographystyle{plain}


\begin{titlepage}
\begin{flushright}
UWThPh-1992-63\\
ESI-1992-1\\
\today
\end{flushright}
\vfill
\begin{center}
{\Large \bf Magnetically Charged Black Holes and Their Stability}\\[40pt]
Peter C. Aichelburg  \\
Institut f\"ur Theoretische Physik \\
Universit\"at Wien \\[5pt]
and \\[5pt]
Piotr Bizon* \\
International Erwin Schr\"odinger Institute for Mathematical Physics \\
A-1090 Vienna, Austria
\vfill
{\bf Abstract} \\
\end{center}

We study magnetically charged black holes in the Einstein-Yang-Mills-Higgs
theory in the limit of infinitely strong coupling of the Higgs field.
Using mixed analytical and numerical methods we give a complete
description of static spherically symmetric black hole solutions, both
abelian and nonabelian. In particular, we find a new class of extremal
nonabelian solutions. We show that all nonabelian solutions are stable
against linear radial perturbations. The implications of our results for
the semiclassical evolution of magnetically charged black holes are
discussed.
\vfill
\noindent *) On leave of absence from Institute of Physics, Jagellonian
University, Cracow, Poland.
\end{titlepage}

\section{Introduction}
It has long been thought that the only static black hole solution in
spontaneously broken gauge theories coupled to gravity
(Einstein-Yang-Mills-Higgs (EYMH) theories) is the Reissner-Nordstr\o m
(RN) solution, with covariantly constant Higgs field and the
electromagnetic field trivially embedded in a nonabelian gauge group
[1]. This belief, based on no-hair properties of black holes, was put
in doubt by the discovery of essentially nonabelian black holes in
EYM theory (so called colored black holes) [2], because it
became clear
that the uniqueness
properties of black holes in Einstein-Maxwell theory [3] are lost when
the gauge field is nonabelian. This suggested that there might also
exist nontrivial (i.e. different than RN) black holes when the Higgs
field is included. Indeed, it was found recently that in the EYMH theory
there are spherically symmetric magnetically charged black holes which
are asymptotically indistinguishable from the RN solution but carry
nontrivial YM and Higgs fields outside the horizon [4,5]. These
essentially nonabelian solutions may be viewed as black holes inside
't~Hooft-Polyakov magnetic monopoles. Because the EYMH field equations are
rather complicated (even for spherical symmetry), one does not have yet
a complete picture of these solutions. In particular, one would like to
know how the solutions depend on external parameters (like coupling
constants)\footnote{This question was examined in the
Prasad-Sommerfield limit in Ref. [5].}, whether there are solutions with
degenerate horizon, and, most important, whether the solutions are
stable.

The aim of this paper is to give answers to these questions within a model
obtained from the EYMH system by taking the limit of infinitely strong
coupling of the Higgs field. In this limit
the Higgs field is effectively ``frozen'' in its vacuum expectation
value. Thus one degree of freedom is reduced which simplifies
considerably the analysis of solutions. This model is interesting on its
own for testing ideas about black holes and as we will argue has the
additional virtue that many of its features carry over to the full
EYMH theory (albeit there are also some important differences).

The paper is organized as follows. In the next section we define the
model and derive the spherically symmetric field equations.

In Section~3 we consider static black hole solutions. The
analysis of asymptotic behaviour of solutions at infinity and at
the horizon allows us to obtain information about possible
global behaviour of solutions. We show that black hole solutions
may exist only if their parameters satisfy certain inequalities.
In particular, we derive necessary conditions for the existence of
extremal black holes.

In Section~4 we describe numerical results and discuss the
qualitative properties of solutions. The bifurcation
structure of
solutions as functions of external parameters is discussed in detail.
We also comment on the status of no-hair conjecture in our model.

Section~5 is devoted to stability analysis. We derive the pulsation
equation governing the evolution of small radial perturbations about
a static solution and solve it numerically. Special emphasis is
placed on the relation between stability properties and the bifurcation
structure of solutions. Using this relation we conjecture about
the stability of black holes in the full EYMH theory.

 Finally, in Section~6 we summarize our results and touch
upon the question of evolution of black holes in the model due
to the
Hawking radiation.

\section{Field Equations}
The Einstein-Yang-Mills-Higgs (EYMH) theory is defined by the action
\begin{equation}
S = \int d^4x \, \sqrt{-g} \left( \frac{1}{16\pi G}R + \cal L_{\rm matter}
\right)
\end{equation}
with
\begin{equation}
{\cal L}_{\rm matter} = - \frac{1}{16\pi} F^2 - \frac{1}{2}(D\phi)^2 -
\lambda(\phi^2 - v^2)^2
\end{equation}
\begin{equation}
F = dA + e[A,A]
\end{equation}
\begin{equation}
D\phi = d\phi + e[A,\phi]
\end{equation}
where the YM connection $A$ and the Higgs field $\phi$ take values in
the Lie algebra of the gauge group $\cal G$. Here we consider ${\cal G}
 = SU(2)$.

There are four dimensional parameters in the theory: Newton's constant
$G$, the gauge coupling constant $e$, the Higgs coupling constant
$\lambda$ and the vacuum expectation value of the Higgs field $v$.
Out of these parameters one can form two scales of length:
$$
\left[ \frac{\sqrt{G}}{e}\right] =
 \left[ \frac{1}{ev}\right] = \mbox{length.}
$$
The ratio of these two scales
\begin{equation}
\alpha = \sqrt{G}\, v
\end{equation}
plays a r\^ole of the effective coupling constant in the model. The
second dimensionless parameter
\begin{equation}
\beta = \frac{\sqrt{\lambda}}{e}
\end{equation}
measures the strength of the Higgs field coupling. In this paper we
consider in detail the case $\beta = \infty$.
In this limit the Higgs field is ``frozen'' in its vacuum
expectation
value $v$.

We are interested in spherically symmetric configurations. It is
convenient to parametrize the metric in the following way
\begin{equation}
ds^2 = - B^2 N dt^2 + N^{-1}dr^2 + r^2(d\vartheta^2 + \sin^2 \vartheta
 d\varphi^2)
\end{equation}
where $B$ and $N$ are functions of $(t,r)$.

We assume that the electric part of the YM field vanishes. Then the
purely magnetic spherically symmetric $su(2)$ YM connection can be
written, in the abelian gauge, as
\begin{equation}
eA = w \tau_1 d\vartheta + (\cot \vartheta \tau_3 + w \tau_2) \sin \vartheta
d\varphi
\end{equation}
where $\tau_i$ ($i = 1,2,3)$ are Pauli matrices and $w$ is a function
of $(t,r)$.

The spherically symmetric Ansatz for the Higgs field in the abelian
gauge is
\begin{equation}
\phi = v \tau_3 h(t,r).
\end{equation}
As we said we concentrate on the limit $\beta = \infty$, which implies
$h(r,t) \equiv 1$. Note that although the Higgs field is constant, its
covariant derivative is nonzero and generates a mass term for the YM
field:
$$
(D\phi)^2 = v^2 w^2.
$$
The YM curvature is given by
\begin{equation}
eF = \dot w dt \wedge \Omega + w' dr \wedge \Omega - (1-w^2) \tau_3 d\vartheta
\wedge \sin \vartheta d\varphi
\end{equation}
where ($\,\dot{} = \partial_t$, ${}' = \partial_r)$ and
$$
\Omega = \tau_1 d\vartheta + \tau_2 \sin \vartheta d\varphi.
$$
{}From the EYMH Lagrangian (1) one derives the following components (in
orthonormal frame) of the stress energy tensor $T_{ab}$
\begin{equation}
\rho \equiv \frac{1}{4\pi} T_{\widehat t \widehat t} = \frac{1}{e^2 r^2}
(Nw'{}^2 + B^{-2} N^{-1} \dot w^2) + \frac{(1-w^2)^2}{2e^2r^4} +
v^2 \frac{w^2}{r^2}
\end{equation}
\begin{equation}
\tau \equiv - \frac{1}{4\pi} T_{\widehat r \widehat r} = - \frac{1}{e^2r^2}
(N w'{}^2 + B^{-2} N^{-1} \dot w^2) + \frac{(1-w^2)^2}{2e^2r^4} +
v^2 \frac{w^2}{r^2}
\end{equation}
\begin{equation}
\sigma \equiv \frac{1}{4\pi} T_{\widehat t \widehat r} = 2 B^{-1} \frac{1}{e^2
r^2}
\dot w w'.
\end{equation}
(In what follows we do not make use of the other nonzero components
$T_{\widehat \vartheta \widehat \vartheta} = T_{\widehat \varphi \widehat
\varphi}$.)

The Einstein equations reduce to the system
\begin{equation}
N' = \frac{1}{r} (1 - N) - 2 G r\rho
\end{equation}
\begin{equation}
B' = Gr B N^{-1} (\rho - \tau)
\end{equation}
\begin{equation}
\dot N = - 2Gr BN\sigma.
\end{equation}
The remaining $\widehat \vartheta \widehat \vartheta$-component
of the Einstein
equations is
equivalent to the YM equation
\begin{equation}
-(B^{-1} N^{-1} \dot w)\dot{} + (BNw')' + \frac{1}{r^2} Bw(1-w^2)
- B e^2 v^2 w = 0.
\end{equation}
For $G = 0$ the equations (14-16) are solved either by the Minkowski
spacetime $N = B = 1$, or by the Schwarzschild spacetime $B=1$,
$N = 1 - 2m/r$, and then the Eq. (17) reduces to the YMH equation on
the given background. For $v = 0$, the system (14 -- 17) reduces to the
EYM equations.

\section{Static Solutions}
In this section we wish to consider static black hole solutions of the
system (14 -- 17). In terms of the dimensionless variable $x = \frac{e}
{\sqrt{G}} r$ the static equations are
\begin{equation}
N' = \frac{1}{x}(1-N) - \frac{2}{x}(Nw'{}^2 + \frac{(1-w^2)^2}{2x^2} +
\alpha^2 w^2)
\end{equation}
\begin{equation}
B' = \frac{2}{x} B w'{}^2
\end{equation}
\begin{equation}
(NBw')' + B\frac{1}{x^2} w(1 - w^2 - \alpha^2 x^2) = 0.
\end{equation}
Note that the function $B$ can be eliminated from Eq. (20) by using
Eq. (19). This system of equations has been studied previously by
Breitenlohner et al. [5] who found asymptotically flat solutions with
naked singularity at $x = 0$. This singularity is an artefact of the
limit $\beta = \infty$, since then there is no symmetry restoration
at the origin and the last term in the expression (11) for the energy
density diverges at $r = 0$. In the black hole case the singularity is
hidden inside the horizon.

We will consider solutions of Eqs. (18 -- 20) in the region
$x \in [x_H,\infty)$, where $x_H$ is the radius of the (outermost)
horizon.
The boundary conditions at $x = x_H$ are
\begin{equation}
N(x_H) = 0, \qquad N'(x_H) \geq 0,
\end{equation}
and the functions $w$, $w'$, $B$ are assumed to be finite.

 At infinity we
impose asymptotic flatness conditions which are ensured by
\begin{equation}
N(\infty) = 1, \qquad w(\infty) = w'(\infty) = 0, \qquad
B(\infty) \;\mbox{ finite.}
\end{equation}
One explicit solution satisfying these boundary conditions is well
known. This is the Reissner-Nordstr\o m (RN) solution
\begin{equation}
w = 0, \qquad B = 1, \qquad
N = 1 - \frac{2m}{x} + \frac{1}{x^2}
\end{equation}
where mass $m \geq 1$ (mass is measured in units $1/\sqrt{G}\,e$). The
corresponding YM curvature is
$$
F = - \tau_3 d\vartheta \wedge \sin \vartheta d\varphi
$$
showing that it is an abelian solution describing a black hole with
 unit magnetic charge (the unit of charge is $1/e$).

It turns out that the system (14-16) admits also nonabelian solutions
which we have
found numerically.
Before discussing the numerical results we want to make some elementary
observations about the global behaviour of solutions satisfying the boundary
conditions (21) and (22). In what follows we assume that $\alpha$ is
nonzero.

First, consider the behaviour of solutions at infinity. The asymptotic
solution of Eqs. (18 -- 20) is
\begin{eqnarray}
w &\simeq& e^{-\alpha x} \nonumber \\
N &\simeq& 1 - \frac{2m}{x} + \frac{1}{x^2} + O(e^{-2\alpha x})
\nonumber \\
B &\simeq& 1 + O(e^{-2\alpha x}) \nonumber
\end{eqnarray}
hence for large $x$ the solutions  are very well approximated by the RN
solution (23).

Second, notice that $w$ has no maxima for $w > 1$ and no
minima for $w < -1$, which follows immediately from Eq. (20). Thus, if
 $w$ once leaves the region $w \in (-1,1)$, it cannot reenter
it. Actually, $w$ stays within this region for all
$x \geq x_H$, because $w(x_H) < 1$ (without loss of generality
we may assume that $w(x_H) > 0$,
because there is a reflection invariance $w \rightarrow -w$). To see this,
consider Eq. (20) at $x = x_H$.
Assuming that $w''(x_H)$ is finite one has
\begin{equation}
\left.N'w' + \frac{1}{x^2} w(1 - w^2 - \alpha^2x^2)\right|_{x=x_H} = 0.
\end{equation}
If $w(x_H) \geq 1$, then $w'(x_H) > 0$, but since there are no maxima
for $w > 1$ the condition $w(\infty) = 0$ cannot be fulfilled.
Thus $w(x_H) < 1$.
Next one can show that $w$ has no positive minima and
negative maxima. Suppose that there is a positive minimum at some
$x_0 > x_H$. Then Eq. (20) implies that the function
$f(x) = 1 - w^2 - \alpha^2x^2$ is negative at $x_0$. For $x \geq x_0$,
$f(x)$ decreases (because $w' \geq 0$), hence there cannot exist a
maximum of $w$ for $x > x_0$, so again the condition $w(\infty) = 0$
cannot be met. One can repeat this argument to show that
$w'(x_H) < 0$, because if $w'(x_H) > 0$, then $f(x_H) < 0$, as
follows from Eq. (20).
Finally, notice that for $x > 1/\alpha$, $w$ cannot have
any extrema because then  $f(x) < 0$, hence there are no positive
maxima and negative minima (whereas other extrema we have already
excluded).

 To summarize, we have shown that if there exits a solution of
Eqs. (18 -- 20), satisfying the boundary conditions (21-22),
then the function $w$ stays in the region
$w \in (-1,1)$ and either monotonically tends to zero or oscillates
around $w = 0$ for $x < 1/\alpha$ and then monotonically goes to zero.

Now, we will show that black hole solutions may exist only if the
parameters $\alpha$ and $x_H$ satisfy certain inequalities. Let
$b \equiv w(x_H)$.
We have shown above that $w'(x_H) < 0$ for $b > 0$, hence from
(24) we obtain
\begin{equation}
\frac{1-b^2}{x^2_H} \geq \alpha^2
\end{equation}
which implies the necessary condition for the existence of a black hole
solution
\begin{equation}
\alpha x_H \leq 1.
\end{equation}
As we shall see in the next section this is not a sufficient condition.

For $x_H < 1$ we can improve the inequality (26). From
Eq. (18) we have
\begin{equation}
N'(x_H) = \frac{1}{x_H} - \frac{(1-b^2)}{x^3_H} - \frac{2\alpha^2b^2}{x_H}
\end{equation}
hence
\begin{equation}
(1-b^2)^2 - x^2_H + 2\alpha^2 x^2_H b^2 \leq 0.
\end{equation}
This inequality has a real solution for $b$ only if the discriminant
\begin{equation}
\frac{1}{4} \Delta = (\alpha^4 x^2_H - 2\alpha^2 + 1)x^2_H
\end{equation}
is nonnegative, which is always satisfied for $x_H \geq 1$ but for
$x_H < 1$ this implies the additional condition
\begin{equation}
\alpha^2 \leq \alpha^2_{\rm max} = \frac{1}{x^2_H}(1 -
\sqrt{1 - x^2_H}).
\end{equation}
It turns out from numerical results that the inequality (30) is also a
sufficient condition for the existence of black holes. Note that when
$x_H \rightarrow 0$ then $\alpha_{\rm max} \rightarrow 1/\sqrt{2}$, which is
the upper
bound for ``regular'' solutions found by Breitenlohner et al. [5].

Finally, let us see whether there may exist extremal black holes in the
model. By extremal we mean a solution with degenerate horizon, i.e.
$N'(x_H) = 0$. For such solutions the inequalities (25) and (28)
are saturated. Eliminating $b^2$ we obtain the condition
\begin{equation}
\alpha^4 x^2_H - 2\alpha^2 + 1 = 0
\end{equation}
which is equivalent to $\Delta = 0$ and is solved
by $\alpha = \alpha_{\rm max}$
given by (30) provided that $x_H \leq 1$.
Let us point out that the necessary conditions for the existence of
extremal black hole solution in EYMH theories were derived some time
ago by Hajicek [6]. In the terminology of Hajicek, Eq. (31) is a
special case of the zeroth-order condition. In the next section we will find
numerically the extremal solutions satisfying (31).

\section{Numerical Results}
The formal power-series expansion of a solution near the horizon is
\begin{eqnarray}
w(x) &=& b + \sum_{k=1}^\infty \frac{1}{k!} w^{(k)}(x_H)(x -
x_H)^k, \nonumber \\
N(x) &=& \sum_{k=1}^\infty \frac{1}{k!} N^{(k)} (x_H)(x - x_H)^k.
\end{eqnarray}
All coefficients in the above series are determined,
through recurrence relations, by $b$, in particular
the expressions for $w'(x_H)$ and $N'(x_H)$ are given by Eqs. (24) and (27).
Thus this expansion (assuming that its radius of convergence is nonzero)
defines a one-parameter family of local
solutions labelled by the initial value $b$.
We use a standard numerical procedure, called shooting method, to
find such values of $b$ for which the local solution extends to a global
solution satisfying the asymptotic boundary conditions (22). Note that
$b$ is not arbitrary but, as follows from Eqs. (25) and (28), must lie
in the interval
\begin{equation}
1 - \alpha^2x^2_H - x_H \sqrt{\alpha^4 x^2_H - 2\alpha^2 +1} \leq b^2
\leq 1 - \alpha^2 x^2_H.
\end{equation}
We find that for every $x_H$, there is a maximal value
 $\alpha_{\rm max}(x_H)$, such that for $\alpha > \alpha_{\rm max}(x_H)$
there are no solutions, while for $\alpha \leq \alpha_{\rm max}$ there
is  exactly one solution for which $w$ is monotonically decreasing.
This solution we call  fundamental, in contrast to solutions
with oscillating $w$ which we call excitations. In what follows
we restrict our attention to fundamental solutions. At the end we
will briefly describe excitations.

The numerical
results fo several values of $x_H$ are summarized in Table 1.
When $x_H \leq 1$, then $\alpha_{\rm max}(x_H)$ is given by Eq. (30).
For $x_H > 1$, $\alpha_{\rm max}(x_H)$ is displayed in Table 2 (in this
case the analytical bound given by Eq. (26) is not sharp).
In Fig. 1 we plot $w(x)$ for
$x_H = 2$ and several values of
$\alpha$.

\paragraph{Table 1:}
Shooting parameter $b$ and mass $m$ (in units $1/\sqrt{G}\,e$)
as functions of $\alpha$ for $x_H = 0.5$,
$x_H = 1$, and $x_H = 2$.
$$
\begin{tabular}{|l|l|l|l|l|l|l|l|l|} \hline
\multicolumn{3}{|c|}{$x_H = 0.5$} & \multicolumn{3}{|c|}{$x_H = 1$} &
\multicolumn{3}{|c|}{$x_H = 2$} \\ \hline
$\alpha$ & $b$ & $m$ & $\alpha$ & $b$ & $m$ & $\alpha$ & $b$ & $m$ \\ \hline
0.01 & 0.99993 & 0.268 & 0.01 & 0.99975 & 0.517 & 0.01 & 0.99907 & 1.018 \\
\hline
0.2 & 0.98296 & 0.572 & 0.1 & 0.98191 & 0.663 & 0.1 & 0.9317 & 1.147 \\
\hline
0.4 & 0.94782 & 0.809 & 0.5 & 0.63099 & 0.9805 & 0.2 & 0.7142 & 1.2286 \\
\hline
0.6 & 0.923028 & 0.954 & 0.74 & 0.1852 & 0.99937 & 0.26 & 0.43074 & 1.248
\\ \hline
0.7 & 0.928359 & 0.990 & 0.9 & 0.00868 & 0.99999 & 0.28 & 0.2414 & 1.2498
 \\ \hline
0.73 & 0.930534 & 0.996 & & & & 0.28867 & 0.012 & 1.25 \\ \hline
\end{tabular}
$$

\paragraph{Table 2:} Maximal value of $\alpha$ as a function of
$x_H$ for $x_H \geq 1$.
$$
\begin{tabular}{|c|c|} \hline
$x_H$ & $\alpha_{\rm max}$ \\ \hline
1 & 1 \\
1.000001 & 0.822 \\
1.01 & 0.744 \\
1.1 & 0.605 \\
1.5 & 0.398 \\
2 & 0.288 \\
10 & 0.055 \\
20 & 0.027 \\ \hline
\end{tabular}
$$

On the basis of analytical and numerical results we have a pretty clear
picture of the qualitative behaviour of solutions. Let us discuss in more
detail how the solutions depend on the parameters $\alpha$ and $x_H$.
First consider the limit $\alpha \rightarrow 0$. In our units this corresponds
to $v \rightarrow 0$. In this limit a solution on any finite region
outside the horizon is well approximated by the Schwarzschild solution
$$
w = 1, \qquad N = 1 - \frac{x_H}{x},
$$
with mass $m = x_H/2$. However, for large
$x$ and small but nonzero $\alpha$ the term $\alpha^2 x^2$ in Eqs. (18)
and (20), becomes dominant and asymptotically the solution tends to the
RN solution.

Next, consider the limit $\alpha \rightarrow \alpha_{\rm max}(x_H)$. In this
case the behaviour of solutions depends on whether $x_H$ is less or
greater than one.
For $x_H \leq 1$, the interval of allowed values of
$b$, given by Eq. (33), shrinks to zero as $\alpha$ goes to
$\alpha_{\rm max}$, and for $\alpha = \alpha_{\rm max}$, $b$ is
uniquely determined by $x_H$. The corresponding limiting solution
describes an {\em extremal} black hole which is essentially
nonabelian.
For $x_H \geq 1$, when $\alpha$ goes to $\alpha_{\rm max}$ the solution
tends to the RN solution and for $\alpha = \alpha_{\rm max}$ coalesce
with it.
Thus for given $x_H \geq 1$, the point $\alpha = \alpha_{\rm max}(x_H)$
is a bifurcation point: for $\alpha > \alpha_{\rm max}$ there is only
one solution (RN), while for $\alpha \leq \alpha_{\rm max}$ the
second (nonabelian) solution appears. The bifurcation diagram in the
plane $(\alpha,x_H)$ is graphed in Fig.~2.
The nonabelian solutions exist only in the region below the curve ABC.
Along the curve AB, the nonabelian solutions are extremal. Along the
curve BC the nonabelian and RN solutions coalesce. The RN solutions
exist for $x_H \geq 1$ and do not depend on $\alpha$ (for $x_H = 1$
the RN solution is extremal).

The bifurcation of solutions is also shown in Fig.~3, where the mass $m$ as
a function of $\alpha$ is graphed for given $x_H \geq 1$.
Notice that the mass of the RN solution $m_{\rm RN} = \frac{1}{2}
(x_H + \frac{1}{x_H})$ is larger than the mass of the nonabelian
solution with the same $x_H$.
This suggests that for $\alpha < \alpha_{\rm max}$, where there are two
distinct solutions with the same radius of the horizon $x_H$, the RN
solution is unstable. We will show in the next section that this is
indeed the case.

Let us comment on the status of the no-hair conjecture in our model. The
no-hair conjecture (in its strong version) states that
stationary black hole
solutions, within a given model, are uniquely determined by global
charges defined as surface integrals at spatial infinity
such as mass, angular momentum and electric or magnetic charge.
In our case, all
solutions are static and have unit
 magnetic charge, so the only global parameter
by which the solutions may differ at infinity is their mass.
For $\alpha \geq 1$ there is only one solution for given mass $m$, namely the
RN solution, hence the strong no-hair conjecture is valid. However,
for $\alpha < 1$ the situation is different. This is illustrated in
Table~3 which shows how the masses of nonabelian and RN solutions depend
on $x_H$ for given $\alpha$.
\paragraph{Table 3:} Masses of nonabelian and RN black holes
as functions of $x_H$ for $\alpha = 0.1$

$$
\begin{tabular}{|c|c|c|} \hline
$x_H$ & $m_{\rm NA}$ & $m_{\rm RN}$ \\ \hline
0.1 & 0.226 & \\
0.2 & 0.274 & \\
1 & 0.663 & 0.75 \\
2 & 1.147 & 1.25 \\
5 & 2.599 & 2.6 \\
5.55 & 2.86508 & 2.86509 \\
5.595 & 2.886865 & 2.886865 \\
6 & & 3.08333 \\
8 & & 4.0625 \\ \hline
\end{tabular}
$$

For the RN solution $x_H \geq 1$, so its mass $m_{\rm RN} = \frac{1}{2}
(x_H + \frac{1}{x_H})$ is bounded from below by 0.75. At
$x_H^{\rm max}(\alpha)$ the nonabelian and RN solutions coalesce, so the
maximal mass of the nonabelian solution is equal to
$m_{\rm RN}(x_H^{\rm max})$. For given $\alpha < 1$ and
$0.75 \leq m < m_{\rm RN}(x_H^{\rm max})$ there are two distinct
(i.e. RN and nonabelian) solutions with the same mass\footnote{For
sufficiently small $\alpha$ the excitations appear so there are even
more than two solutions with the same mass.}. This violates the
strong no-hair conjecture.

We will show in the next section that in the region where two distinct
solutions coexist, only one of them is stable. Thus, in our model, a
{\em weak} no-hair conjecture holds, i.e. global charges determine
uniquely the {\em stable} black hole solution. However, as we will
argue, in the full EYMH theory even the weak no-hair conjecture is
violated.

Finally, let us briefly consider solutions for which $w$ is not
monotonic. Such solutions may be viewed as excitations of the
fundamental solutions described above. Their existence
in the EYMH model was first noticed in Ref. 5. To understand the
existence of excitations it is useful to consider the limit $\alpha
\rightarrow 0$. Then, the Eqs. (18 -- 20) become the EYM equations, which have
a
countable family of so-called colored black holes [2].
These solutions are labelled by an integer $n =$ number of nodes of $w$,
and in contrast to the $\alpha \neq 0$ case, they have zero magnetic charge
(because $w(\infty) = \pm 1$). For sufficiently small $\alpha$, the
excitations may be viewed as singular perturbations (in $\alpha$) of
colored black holes.
As long as $\alpha x \ll 1$, the excitations are well approximated by
colored black hole solutions (with the same number of nodes for $w$),
but for large $x$ the term $\alpha^2 x^2$ in Eqs. (18) and (19) becomes
dominant and forces $w$ to decay exponentially. This is shown in
Fig.~5, where the $n=1$ excitation is graphed for two values of $\alpha$
and compared to the $n=1$ colored black hole solution.

It is rather difficult to find numerically the excitations with large
$n$, however we expect that for given $x_H$ and $\alpha \neq 0$, the number
of excitations is finite (in contrast to the $\alpha = 0$ case, where
there are infinitely many solutions).
Our expectation is based on the monotonicity of solutions for $x < 1/\alpha$,
proven in Section~3. The existence of excitations with arbitrarily large
$n$ would be inconsistent with this property, because, in analogy to the
colored black holes, the location of nodes is expected to extend to
infinity as $n$ increases.

The bifurcation structure of excitations is similar to that for the
fundamental solution. For given $x_H \geq 1$ there is a decreasing
sequence $\{ \alpha^0_{\rm max},\alpha^1_{\rm max},\ldots\}$ of bifurcation
points such that at $\alpha^n_{\rm max}$ the $n$-th excitation
bifurcates from the RN branch (by zeroth excitation we mean the
fundamental solution). As we will discuss in the next section this
picture is closely related to the stability properties of the RN
solution.

\section{Stability Analysis}
In this section we address the issue of linear stability of the black
hole solutions described above. To that purpose we have to study the
evolution of linear perturbations about the equilibrium configurations.
Since we do not expect nonspherical instability (and since,
admittedly, the analysis of nonspherical perturbations would be
extremely complicated), we restrict our study to radial perturbations.
For radial perturbation the stability analysis is relatively simple,
because the spherically symmetric gravitational field has no dynamical
degrees of freedom and therefore the perturbations of metric
coefficients are determined by the perturbations of matter fields.
This was explicitly demonstrated for the EYM system by Straumann and Zhou
 [7]
 who derived the pulsation equations governing the evolution
of radial normal modes of the YM field.
Below we repeat their derivation with a slight modification due to
the presence of the mass term in Eqs. (14 -- 17).

We define the functions $a$ and $b$ by $e^a \equiv BN$ and $e^b \equiv N$
and write the perturbed fields as
\begin{eqnarray}
w(r,t) &=& w_0(r) + \delta w(r,t) \nonumber \\
a(r,t) &=& a_0(r) + \delta a(r,t) \\
b(r,t) &=& b_0(r) + \delta b(r,t) \nonumber
\end{eqnarray}
where $(w_0,a_0,b_0)$ is a static solution. We insert the expressions (34)
into Eqs. (14 -- 17) and keep first order terms in the
perturbations. Hereafter we use dimensionless variables
$$
\tau = \frac{e}{\sqrt{G}}\: t \qquad \mbox{and} \qquad
x = \frac{e}{\sqrt{G}}\: r
$$
and we omit the subscript 0 for static solutions. From (16) we obtain
\begin{equation}
\delta \dot b = - \frac{4}{x} w' \delta \dot w.
\end{equation}
The asymptotically flat solution of this equation is
\begin{equation}
\delta b = - \frac{4}{x} w' \delta w.
\end{equation}
Linearization of Eq. (15) yields
\begin{equation}
\delta a' - \delta b' = \frac{4}{x} w' \delta w' .
\end{equation}
Thus, using (36), we obtain
\begin{equation}
\delta a' = \frac{4}{x^2} w' \delta w - \frac{4}{x} w'' \delta w .
\end{equation}

Multiplying Eq. (17) by $e^{-a}$ and linearizing we get
\begin{equation}
- e^{-2a} \delta \ddot w + e^{-a}(e^a \delta w')' + \delta a'w'
- \delta b e^{-b} w \left[ \frac{1}{x^2} (1-w^2) - \alpha^2\right] +
e^{-b} \left[ \frac{1}{x^2}(1 - 3w^2) - \alpha^2\right] \delta w = 0.
\end{equation}
Now, we insert (36) and (38) into (39) and make the Ansatz
\begin{equation}
\delta w = e^{i\omega\tau} \xi(x)
\end{equation}
to obtain the eigenmode equation
\begin{equation}
- e^a(e^a \xi')' + U \xi = {\omega}^2 \xi
\end{equation}
where
\begin{equation}
e^{-a} U = - \frac{4}{x^2} e^a(1+a'x)w'{}^2 - \frac{8}{x} e^{a-b} ww'
\left(\frac{1-w^2}{x^2} - \alpha^2\right) - e^{a-b}
\left[ \frac{1}{x^2} (1-3w^2) - \alpha^2\right].
\end{equation}
The potential $U$ is a smooth bounded function which vanishes at $x_H$
and tends to $\alpha^2$ for $x \rightarrow \infty$.

One can introduce the tortoise radial coordinate to transform Eq. (41)
into the one-dimensional Schr\"odinger equation. However we will
not
do so, because the numerical analysis is easier when one uses the
coordinate $x$.

A static solution $(w,a,b)$ is stable if there are no integrable
eigenmodes $\xi$ with negative ${\omega}^2$. To check this we have applied a
(slightly modified)
rule of nodes for Sturm-Liouville systems [8], which states that the
number of negative eigenmodes is equal to the number of nodes of the
zero eigenvalue solution.
Namely we have considered Eq. (41) with negative ${\omega}^2$ and looked
how the solution satisfying $\xi(x_H) = 0$ and $\xi'(x_H) > 0$
behaves as ${\omega}^2 \rightarrow 0^-$.
We have found that the function $\xi$ has no nodes, when $U$ is
determined by the fundamental nonabelian solution
 (for all allowed values of the
parameters). Thus we conclude that the fundamental nonabelian solutions are
stable. When $U$ is determined by the $n$-th excitation the function
$\xi$ seems to have exactly $n$ nodes (we have checked this up
to $n=2$), hence the $n$-th excitation has $n$ unstable modes.
 For $\omega^2 = 0$ and $\alpha \rightarrow \alpha^n_{\rm max}$
the
function $\xi$, after making $n$ oscillations,
 tends to zero at infinity which signals the existence
of static perturbations (bifurcation points).

One can apply the same method to examine the stability of the RN
solution. This was done by Lee et al. [9] in the full EYMH theory.
They showed that the pulsation equation for the Higgs field (which in
the case of RN solution decouples from the pulsation equation for the
YM field) has no unstable modes. Thus the question of stability of the
RN solution does not depend on the parameter $\beta$ (strength of the
Higgs coupling) and reduces for every $\beta$ to the Eq. (41), where
now
\begin{equation}
U = e^{a-b} \frac{\alpha^2 x^2 -1}{x^2}.
\end{equation}
For given $x_H \geq 1$, the existence of negative eigenmodes in this
potential depends on $\alpha$.
For $\alpha > \alpha_{\rm max}(x_H)$ there are no negative eigenvalues,
hence the RN solution is stable. For $\alpha < \alpha_{\rm max}(x_H)$,
the RN is unstable. Equivalently, one can say that for given $\alpha$, the
RN is stable if $x_H > x_H^{\rm max}(\alpha)$, otherwise it is
unstable.

We have found numerically that when $\alpha$ decreases the RN solution
picks up additional unstable modes. That is, there is a decreasing
sequence $\{ \alpha_n\}$, where $\alpha_0 = \alpha_{\rm max}$, such that
in the interval $\alpha_{n+1} < \alpha < \alpha_n$ there are exactly
$n$ unstable modes (we have checked this up to $n = 4$). This is
consistent with the fact that the RN solution has infinitely many
unstable modes for $\alpha = 0$, as follows easily from (43). We are
convinced that the sequence $\{ \alpha_n\}$ coincides with the sequence
$\{ \alpha^n_{\rm max}\}$ of the bifurcation points at which the $n$-th
excitation appears. Due to highly unstable behaviour of solutions near
$\alpha^n_{\rm max}$ we were able to verify this assertion with sufficient
numerical accuracy only up to $n = 1$. However, as long as there are no
other bifurcation points (as we believe), our conclusion is
based on the
general theorem of the bifurcation theory which says that
 if the operator governing small fluctuations if self-adjoint
(hence its eigenvalues are real), then at
the bifurcation point one eigenvalue passes through zero, whereas
elsewhere the eigenvalues cannot change sign [10].

One could have anticipated the stability properties of solutions from a
mere comparison of masses. In the region of the plane $(\alpha,x_H)$
where two distinct solutions coexist (with the same $x_H$), the solution with
lower mass (i.e. nonabelian) is stable, while the solution with higher
mass (i.e. RN) is unstable. However, it should be emphasized that in more
complicated situations the naive comparison of masses is not conclusive
for stability. Actually, this happens in the full EYMH theory.
To see this, let us consider the bifurcation diagram $(m,\alpha)$ for
$x_H > 1$ in the full EYMH theory. As follows from the results of Ref. 5
in the Prasad-Sommerfield limit and our preliminary results for finite
$\beta$, this diagram is more complicated than that shown in Fig.~3. We
sketch it in Fig.~5.

The horizontal line in Fig.~5 represents the mass of the RN solution.
There are two branches of (fundamental) nonabelian solutions: the
upper branch AB and the lower branch CB. These two branches merge at the
bifurcation point B, which corresponds to $\alpha_{\rm max}(x_H)$.
There is a second bifurcation point A, corresponding to some
$\alpha_0(x_H)$, where the upper branch merges with the RN solution.
One can use these facts to infer the stability properties of solutions.
As we have pointed out above the existence of a bifurcation point is a
necessary condition for the transition between stability and instability.
However, it is not a sufficient condition, since it is only when the
lowest eigenvalue passes through zero that the stability changes.
Therefore, if there are two branches which merge at the bifurcation
point and one of them is stable then the other one will have exactly one
unstable mode.
Applying this reasoning in the present context, we see that if the lower
branch is stable for small $\alpha$ (what we expect), then the whole lower
branch must be stable, whereas the upper one is unstable with exactly one
negative mode. By a similar argument the RN solution becomes unstable
at $\alpha_0$. Therefore, in the region between $\alpha_0$ and
$\alpha_{\rm max}$ there are two stable solutions and one unstable.
Note that in this region there is no relation between stability and mass.
In particular the point in Fig.~5 at which the lower branch crosses the
horizontal line is not related to change in stability (contrary to
the suggestion in Ref. 5).

The analogous bistability region exists in the plane $(m,x_H)$ for
fixed $\alpha$. This means that for given mass $m$ there are two
distinct stable black hole solutions which violates the weak no-hair
conjecture [11].

\section{Discussion}
We have analyzed static spherically black hole solutions in the strong
Higgs coupling limit of the
 EYMH theory. We have found that the spectrum
of solutions depends on the value of the dimensionless parameter
$\alpha$ characterizing the model.
Our results may be summarized as follows.

 For $\alpha > 1$ there is only
one branch of solutions, namely the RN family (parametrized by the radius
of the horizon $r_H \geq \sqrt{G}/e$). For $\alpha < 1$ there appears
a second (fundamental) branch of essentially nonabelian
solutions. The radius of the horizon of
these solutions is confined to the interval
$r^{\rm min}_H(\alpha) \leq r_H \leq r^{\rm max}_H(\alpha)$. The upper
bound $r^{\rm max}_H(\alpha)$ is a bifurcation point at which the two
branches merge. The lower bound $r^{\rm min}_H(\alpha)$ is nonzero only
for $1/\sqrt{2} < \alpha < 1$. The solutions with
$r_H = r^{\rm min}_H(\alpha) > 0$ are extremal in the sense that they
have a degenerate event horizon and therefore
their Hawking temperature vanishes. The analysis of linear radial
perturbations about the solutions shows that the nonabelian solutions
are stable whereas the RN solution changes stability at the bifurcation
point, i.e. it is stable for $r_H > r^{\rm max}_H(\alpha)$ and unstable
for $r_H < r^{\rm max}_H(\alpha)$. For sufficiently small
 values of $\alpha$ there
exist also other branches of nonabelian solutions, which may be viewed as
excitations of the fundamental solution. All excitations are unstable.

These results have interesting implications for the fate of evaporating
black holes in our model. Consider
 the RN black hole with unit magnetic
charge and large mass. There are three different scenarios of its evolution
due to the Hawking radiation
depending on the value of $\alpha$. For $\alpha > 1$ the situation is the
same as in the Einstein-Maxwell theory. The RN black hole emits thermal
radiation and loses mass. When the horizon shrinks to
$r_H = \sqrt{G}/e$ (extremal limit), the temperature drops to zero and
the evaporation stops.

For $\alpha < 1$, the RN black hole contracts to $r_H = r_H^{\rm max}
(\alpha)$ where it becomes classically unstable. The further evolution
proceeds along the classically stable nonabelian
 branch and depends on whether $\alpha$
is smaller or greater than $1/\sqrt{2}$. For $\alpha > 1/\sqrt{2}$, the
temperature drops to zero when the horizon shrinks to
$r_H^{\rm min}(\alpha)$ (extremal limit) and the black hole settles
down as an {\em extremal nonabelian solution}.
For $\alpha < 1/\sqrt{2}$, the temperature grows as the black hole contracts,
so the black hole evaporates completely leaving behind a magnetic monopole
remnant.
This last scenario was first suggested by Lee et al. [6] as a possible
consequence of the instability of RN solution.

\section*{Acknowledgements}
P.B. would like to acknowledge the hospitality of the Aspen Center of
Physics where part of this work was carried out. This work was supported
in part by
the Fundaci\'on Federico.

\section*{Figure captions}
\begin{description}
\item[Fig.1] The solution $w(x)$ for $x_H=2$ and $\alpha=0.01$
(solid line), $\alpha=0.2$ (dashed line), and $\alpha=0.288$
(dotted line).

\item[Fig.2] The bifurcation diagram in the $(\alpha, x_H)$ plane.
Excitations are not included.
\item[Fig.3] Mass $m$ (in units $1/\sqrt{G}\,e$) of the
fundamental nonabelian solution as a
function of $\alpha$ for $x_H=2$. The
mass of the RN solution is $m_{RN}=1.25$.

\item[Fig.4] The $n=1$ excitations for $x_H=1$. The dashed line represents
the $n=1$ colored black hole solution.

\item[Fig.5] The sketch of the bifurcation structure of black
hole solutions with $x_H \geq 1$ in full EYMH theory.
Excitations are not included.

\end{description}

\newpage


\newpage
\begin{thebibliography}{99}
\bibitem{1} F. A. Bais and R. J. Russel, Phys. Rev. {\em D 11}
(1975) 449; \\
Y. M. Cho and P. G. O. Freund, Phys. Rev. {\em D 12} (1975) 1588.
\bibitem{2} R. Bartnik and J. Mckinnon, Phys. Rev. Lett. {\em
61} (1988) 41;\\
P. Bizon, Phys. Rev. Lett. {\em 64} (1990) 2844;\\
H. P. K\"unzle and A. K. M. Masoud-ul-Alam, J. Math. Phys. {\em
31} (1990) 928;\\
M. S. Volkov and D. V. Gal'tsov, Sov. J. Nucl. Phys. {\em 51}
(1990) 1171.
\bibitem{3} For a review, see P. O. Mazur, in
{\em Proceedings of the 11th International Conference on General
Relativity and Gravitation}, ed. M. A. H. MacCallum (Cambridge University
Press, 1987).

\bibitem{4} K. Lee et al., Phys. Rev. {\em D 45} (1992) 2751.
\bibitem{5} P. Breitenlohner et al, Nucl. Phys. {\em B383} (1992) 357.
\bibitem{6} P. Hajicek, Proc. Roy. Soc. {\em A 386} (1983) 223;
J. Phys. {\em A 16} (1983) 1191.
\bibitem{7} N. Straumann and Z.-H. Zhou, Phys. Lett {\em B 237}
(1990) 353; Phys. Lett. {\em B 241} (1990) 33.
\bibitem{8} M. Reed and B. Simon, {\em Methods of Modern
Mathematical Physics}, vol. IV (Academic Press, New York, 1978).
\bibitem{9} K. Lee et al., Phys. Rev. Lett. {\em 68} (1992) 1100.
\bibitem{10} S. N. Chow, J. K. Hale, {\em Methods of Bifurcation Theory}
(Springer, Berlin, 1982).
\bibitem{11} To our knowledge, the violation of the weak no-hair
conjecture was first observed in the Einstein-Skyrme model by
P. Bizon and T. Chmaj, {\em Gravitating Skyrmions},
University of Vienna preprint, UWThPh-1992-23,
to be published in Phys. Lett. {\em B}.


\end{thebibliography}
\end{document}